\documentclass[a4paper,fleqn,usenatbib,letters]{mnras}

\pdfoutput=1

\usepackage{amssymb,amsmath,latexsym,mathrsfs}
\usepackage{graphicx,subfigure}
\usepackage{epsfig}
\usepackage{varioref,xr-hyper}
\usepackage{color}
\usepackage{multirow}
\usepackage{array}
\usepackage{hyperref}
\usepackage{wasysym}
\usepackage{color}
\usepackage{float}
\usepackage{mathtools}
\usepackage[utf8]{inputenc}
\usepackage[T1]{fontenc}

\title[Non-parametric spatial curvature inference using late-universe cosmological probes]{Non-parametric spatial curvature inference using late-universe cosmological probes}

\author[S. Dhawan, J. Alsing \& S. Vagnozzi]{
Suhail Dhawan,$^{1}$\thanks{E-mail: \href{mailto:suhail.dhawan@ast.cam.ac.uk}{suhail.dhawan@ast.cam.ac.uk} (SD)}
Justin Alsing$^{2,3}$\thanks{E-mail: \href{mailto:justin.alsing@fysik.su.se}{justin.alsing@fysik.su.se} (JA)}
and Sunny Vagnozzi$^{1}$\thanks{E-mail: \href{mailto:sunny.vagnozzi@ast.cam.ac.uk}{sunny.vagnozzi@ast.cam.ac.uk} (SV)}
\\
$^{1}$Kavli Institute for Cosmology and Institute of Astronomy, University of Cambridge, Madingley Road, Cambridge CB3 0HA, UK
\\
$^{2}$The Oskar Klein Centre for Cosmoparticle Physics, Department of Physics, Stockholm University, SE-10691 Stockholm, Sweden\\
$^{3}$Imperial Centre for Inference and Cosmology, Astrophysics Group,
Imperial College London, Blackett Laboratory, Prince Consort Road, \\ London SW7 2AZ, UK
}

\date{Accepted XXX. Received YYY; in original form ZZZ}

\pubyear{2021}

\begin{document}
\label{firstpage}
\pagerange{\pageref{firstpage}--\pageref{lastpage}}
\maketitle

\begin{abstract}
Inferring high-fidelity constraints on the spatial curvature parameter, $\Omega_{\rm K}$, under as few assumptions as possible, is of fundamental importance in cosmology. We propose a method to non-parametrically infer $\Omega_{\rm K}$ from late-Universe probes alone. Using Gaussian Processes (GP) to reconstruct the expansion history, we combine Cosmic Chronometers (CC) and Type Ia Supernovae (SNe~Ia) data to infer constraints on curvature, marginalized over the expansion history, calibration of the CC and SNe~Ia data, and the GP hyper-parameters. The obtained constraints on $\Omega_{\rm K}$ are free from parametric model assumptions for the expansion history, and are insensitive to the overall calibration of both the CC and SNe~Ia data (being sensitive only to relative distances and expansion rates). Applying this method to \textit{\textit{Pantheon}} SNe~Ia and the latest compilation of CCs, we find $\Omega_{\rm K} = -0.03 \pm 0.26$, consistent with spatial flatness at the $\mathcal{O}(10^{-1})$ level, and independent of any early-Universe probes. Applying our methodology to future Baryon Acoustic Oscillations and SNe~Ia data from upcoming Stage IV surveys, we forecast the ability to constrain $\Omega_{\rm K}$ at the $\mathcal{O}(10^{-2})$ level.
\end{abstract}


\section{Introduction}
\label{sec:intro}

Measuring the spatial curvature of the Universe is of long-standing interest in cosmology. Determining the sign and value of the curvature parameter $\Omega_{\rm K}$ would be of great significance for fundamental physics, given its implications for both early- and late-Universe physics in relation to the inflationary paradigm and the ultimate fate of the Universe respectively. This question has received significant, renewed interest in the past years, particularly in light of the \textit{Planck} Cosmic Microwave Background (CMB) legacy data release~\citep{Aghanim:2018eyx}, whose temperature and polarization data, taken on their own, could appear at face value to suggest that the Universe might not be spatially flat.

The ability of primary CMB data to constrain $\Omega_{\rm K}$ is, ultimately, limited by the geometrical degeneracy~\citep{Bond:1997wr,Efstathiou:1998xx,Zaldarriaga:1997ch}, which can be broken by including ``external'' data, such as Baryon Acoustic Oscillation (BAO) measurements. While this usually results in the inference of the Universe being spatially flat to within $\approx 1\sigma$~\citep[see e.g.][]{Alam:2016hwk,Aghanim:2018eyx,Efstathiou:2020wem,Vagnozzi:2020zrh,Vagnozzi:2020dfn}, doubts have been cast on the soundness of such combinations, suggesting the possibility of there being a ``curvature tension'' in current data~\citep[see e.g.][]{Handley:2019tkm,DiValentino:2019qzk,DiValentino:2020hov}.

Without going into the details of the previous discussions, it is clearly of considerable interest to obtain high-fidelity measurements of $\Omega_{\rm K}$.~\footnote{For earlier works examining constraints and forecasted constraints on $\Omega_{\rm K}$, see for instance~\cite{Vardanyan:2009ft,Takada:2015mma,Leonard:2016evk,Yu:2016gmd,Witzemann:2017lhi,Denissenya:2018zcv,Li:2019qic,Park:2019emi,Khadka:2020hvb,Nunes:2020uex,Benisty:2020otr,2021arXiv210108817C}.} Moreover, many of the constraints on $\Omega_{\rm K}$ available in the literature depend on the assumed parametric form of the late-time expansion rate. It is therefore also desirable to obtain non-parametric, purely geometrical constraints on $\Omega_{\rm K}$ from late-Universe cosmological probes alone, independent of and complementary to the early-Universe CMB constraints.

In this \textit{Letter}, we propose a non-parametric approach for inferring $\Omega_{\rm K}$ using only late-Universe distance and expansion rate indicators, in the form of Type Ia Supernovae (SNe~Ia) and cosmic chronometers (CC). Imposing a Gaussian Process (GP) smoothness prior on the expansion history, we use the CC and SNe~Ia data to jointly infer the curvature parameter, (non-parametric) expansion history, calibration of the CC and SNe~Ia data, and GP hyper-parameters. The resulting marginal constraints on $\Omega_{\rm K}$ are free from parametric model assumptions about the expansion history. Moreover, being only sensitive to relative distances and expansion rates from the SNe~Ia and CC measurements respectively, our constraints on $\Omega_{\rm K}$ are insensitive to the overall calibration (and associated systematics) in either dataset. Applying our method to current SNe~Ia data from the \textit{Pantheon} compilation \citep{2018ApJ...859..101S} and the latest compilation of CC measurements, we find $\Omega_{\rm K} = -0.03 \pm 0.26$, consistent with flatness at the $\sim \mathcal{O}(10^{-1})$ level. Applying our methodology to future Stage-IV dark energy missions, we forecast that these data will be able to deliver $\sim \mathcal{O}(10^{-2})$-level constraints on $\Omega_{\rm K}$, competitive with the current CMB-only constraints from \textit{Planck}~\citep{Aghanim:2018eyx} and ACT+WMAP~\citep{Aiola:2020azj}.

\section{Data and Methodology}
\label{sec:data}
Over the past two decades, Type Ia Supernovae (SNe~Ia) and cosmic chronometers (CC) have emerged as important probes of the late-time expansion history. SNe~Ia are excellent late-time distance indicators, and have been used among the other things to infer the existence of cosmic acceleration, to measure the properties of the dark energy (DE) component responsible for cosmic acceleration~\citep{Riess:1998cb,Perlmutter:1998np,2018ApJ...859..101S}, and (once calibrated with independent distances to their host galaxies) to infer the present day value of the expansion rate, \textit{i.e.} the Hubble constant $H_0$~\citep{Riess:2019cxk,2019ApJ...882...34F}. SNe~Ia are sensitive to the luminosity distance $d_\mathrm{L}(z)$ via the distance modulus:
\begin{equation}
\mu=5\log(d_\mathrm{L})+25\,,
\label{eq:mu}
\end{equation}
where the luminosity distance is given by:
\begin{equation}
d_\mathrm{L} = \frac{cz}{H_0 \sqrt{\vert \Omega_{\rm K}\vert}} \mathrm{sinn} \left \{ \sqrt{|\Omega_{\rm K}|}\int_0^z \frac{dz^{'}}{E(z^{'})} \right \}\,,
\label{eq:dl}
\end{equation}
with sinn indicating either $\sin$ or $\sinh$ depending on whether $\Omega_{\rm K}<0$ or $\Omega_{\rm K}>0$, and where $H(z)$ is the Hubble expansion rate, $E(z) \equiv H(z)/H_0$ is the normalized expansion rate, and $M$ is the absolute SNe~Ia calibration. While an absolute SNe~Ia luminosity calibration (achieved for instance through Cepheids or the Tip of the Red Giant Branch) is necessary to determine \textit{absolute} distances and hence $H_0$, for the purpose of inferring $\Omega_{\rm K}$ only \textit{relative} distances are required. Therefore, inferences on the curvature parameter are expected to be insensitive to the overall SNe~Ia calibration, as we will explicitly demonstrate later.

CCs are tracers of the evolution of the differential age of the Universe as a function of redshift, from which the Hubble expansion rate $H(z)$ can be inferred directly, essentially by inverting the age-redshift relation~\citep{Jimenez:2001gg}. Massive, early, passively-evolving galaxies have been found to be very good tracers in this sense~\citep[see e.g.][for important works in this direction]{Cimatti:2004gq,Thomas:2009wg,Moresco:2015cya,Moresco:2018xdr,Moresco:2020fbm}, and have been used extensively over the past two decades to measure $H(z)$ up to $z \approx 2$~\citep[see e.g.][whose measurements we use]{Jimenez:2003iv,Simon:2004tf,Stern:2009ep,Moresco:2012by,Moresco:2015cya,Moresco:2016mzx,Ratsimbazafy:2017vga}. When combining direct measurements of $H(z)$ from CCs with SNe~Ia to constrain curvature, the information added by CCs only influences our $\Omega_{\rm K}$ inference via $E(z) = H(z)/H_0$. Therefore, as for SNe~Ia, inferences on the curvature parameter are expected to be insensitive to the absolute calibration of the CC measurements. 

In this work we combine SNe~Ia and CC data to constrain the spatial curvature parameter $\Omega_\mathrm{K}$, assuming a non-parametric model for the expansion history $H(z)$, i.e. independent of any fixed parametric cosmological model for the expansion rate. In this way, we are able to infer constraints on spatial curvature that are independent of \textit{a)} any assumed cosmological model for the late-time expansion, \textit{b)} the absolute calibration of either the SNe~Ia or CC measurements, and \textit{c)} early-Universe measurements.
\subsection{Data and likelihoods}
We use the latest compilation of SNe~Ia distance moduli measurements from the \textit{Pantheon} compilation~\citep{2018ApJ...859..101S}, combined with 31 CC $H(z)$ measurements in the range $0.07<z<1.965$. The CC measurements have been compiled in~\cite{Jimenez:2003iv,Simon:2004tf,Stern:2009ep,Moresco:2012by,Moresco:2015cya,Moresco:2016mzx,Ratsimbazafy:2017vga}, and are summarized in Tab.~I of~\cite{Vagnozzi:2020dfn}.

The SNe~Ia data comprise measurements of the distance moduli as a function of redshift, $\mu(z)$. We take the binned \textit{Pantheon} data vector $\mathbf{d}_\mathrm{SN}$ and systematics-marginalized covariance matrix $\mathbf{C}_\mathrm{SN}$ from~\cite{2018ApJ...859..101S}, with log-likelihood given by (up to an additive constant):
\begin{align}
    \mathrm{ln}\,P(&\mathbf{d}_\mathrm{SN} | \Omega_{\mathrm{K}}, H(z), M) = \nonumber \\
&-\frac{1}{2}\left[\hat{\boldsymbol{\mu}} - \boldsymbol{\mu}(\Omega_{\mathrm{K}},\, H(z),\, M)\right]^\mathrm{T}\mathbf{C}_\mathrm{SN}^{-1} \left [ \hat{\boldsymbol{\mu}} - \boldsymbol{\mu}(\Omega_{\mathrm{K}},\, H(z),\, M) \right ] \,,
\label{eq:loglklsne}
\end{align}
where $\hat{\boldsymbol{\mu}}$ is the data vector of measured distance moduli, and $\mu_i = \mu(z_i,\, \Omega_{\rm K},\, H(z),\, M)$ are the predicted distance moduli for redshift bins $\{z_i\}$, see Eqs.~(\ref{eq:mu},\ref{eq:dl}).

For the CC data, with data vector given by $\mathbf{d}_\mathrm{CC}$, we assume independent Gaussian uncertainties, and hence a log-likelihood given by (up to an additive constant):
\begin{align}
    \mathrm{ln}\,P(\mathbf{d}_\mathrm{CC} | H(z)) = -\frac{1}{2} \sum_j (\hat{H}_j - H(z_j))^2/\sigma_\mathrm{CC,\,j}^2\,,
\end{align}
where the CC data consists of measurements of $\{\hat{H}_j\}$ at redshifts $\{z_j\}$, with uncertainties ${\sigma_\mathrm{CC,\,j}}$ respectively.
\subsection{Priors}
We set (improper) uniform priors on the spatial curvature parameter $\Omega_\mathrm{K}$ and the absolute SNe~Ia calibration $M$. With regards to the expansion history $H(z)$, we proceed non-parametrically, assuming solely that $H(z)$ is a smooth function of $z$. The ``smoothness'' of $H(z)$ is controlled by a set of hyper-parameters $\boldsymbol\eta$, which are included as free parameters to be eventually marginalized over. To this end, we impose a Gaussian process (GP) prior on $H(z)$:
\begin{align}
P(H(z) | \boldsymbol\eta) = \mathcal{N}(H(z) | m(z), K_{\boldsymbol\eta})\,,
\label{eq:gpprior}
\end{align}
where $m(z)$ is the prior mean. On the other hand, the prior covariance between the values of $H(z)$ at two redshifts $z$ and $z'$ is specified by the kernel function $K_{\boldsymbol\eta}$: 
\begin{align}
K_{\boldsymbol\eta} \equiv K_{\boldsymbol\eta}(z, z') = \langle(H(z) - m(z))(H(z') - m(z'))\rangle\,.
\label{eq:kernel}
\end{align}
The properties of the chosen kernel function (governed by its functional form and hyper-parameters $\boldsymbol\eta$) characterize the prior assumptions on the smoothness of the expansion history as a function of redshift.

In this work, we adopt two common choices for the Gaussian process kernel functions. As our baseline, we use the squared-exponential kernel, given by:
\begin{eqnarray}
K_{\boldsymbol\eta}(z, z') = a^2\exp \left ( -\frac{|z - z'|^2}{2\ell^2} \right )\,,
\end{eqnarray}
where the hyper-parameters $\boldsymbol\eta \equiv (a,\ell)$ control the amplitude and length-scale of the prior covariance respectively. To test the sensitivity of our results to the choice of GP kernel, we also consider the Mat\'{e}rn-$3/2$ kernel, given by:
\begin{eqnarray}
K_{\boldsymbol\eta}(z, z') = a^2\left(1 + \frac{\sqrt{3|z-z'|^2}}{\ell}\right)\mathrm{exp}\left(-\frac{\sqrt{3|z-z'|^2}}{\ell}\right)\,,
\end{eqnarray}
where again the hyper-parameters $\boldsymbol\eta \equiv (a,\ell)$ control the amplitude and length-scale of the prior covariance. We set flat (positive) priors on the kernel hyper-parameters for both the squared-exponential and Mat\'{e}rn-$3/2$ kernels, and take the GP prior mean to be $m(z) = 100$ (setting an appropriate scale for $H(z)$ over the redshift range of interest).
\subsection{Joint posterior and inference strategy}
In practice, in order to sample from the joint posterior for $\Omega_\mathrm{K}$, $M$ and $H(z)$ including the unknown ``function'' $H(z)$, one has to discretize $H(z)\rightarrow \mathbf{H}$ at some redshifts $\{z_k\}$, and include those nodes as free parameters (to be marginalized over).~\footnote{Note that in many common use cases for Gaussian processes under Gaussian likelihoods, the latent function can be marginalized over analytically. However, in this case since the SNe~Ia likelihood is non-Gaussian in $H(z)$, the function needs to be discretized and explicitly sampled over in the inference pipeline.} The Gaussian process prior on $H(z)$ translates into a simple multivariate Gaussian prior on $\mathbf{H}$:
\begin{eqnarray}
P(\mathbf{H} | \boldsymbol\eta) = \frac{1}{\sqrt{|2\pi\mathbf{K}_{\boldsymbol\eta}|}}\mathrm{exp}\left[{-\frac{1}{2}(\mathbf{H} - \mathbf{m})^\mathrm{T}\mathbf{K}_{\boldsymbol\eta}^{-1}(\mathbf{H} - \mathbf{m})}\right]
\end{eqnarray}
with mean $m_k = m(z_k)$ and covariance (specified by the Gaussian process kernel function) $K_{\boldsymbol\eta,\,kl} = K_{\boldsymbol\eta}(z_k, z_l)$. We choose redshift nodes for $H(z)$ that are dense enough in redshift so that the luminosity distance integrals [Eq. \eqref{eq:dl}] in the SNe~Ia likelihood can be performed numerically,~\footnote{This ensures that the resulting distance integrals are accurate enough, i.e. with errors much smaller than the (root) diagonals of the SNe~Ia data covariance.} and include additional nodes at the CC redshift values (as required for evaluating the CC likelihood).

The full joint posterior over $\Omega_\mathrm{K}$, $\mathbf{H}$, $M$, and kernel hyper-parameters $\boldsymbol{\eta}$, is therefore given by:
\begin{align}
    P(\Omega_\mathrm{K}, \mathbf{H}, M, \boldsymbol\eta) \propto\; &P(\mathbf{d}_\mathrm{SN} | \Omega_\mathrm{K}, \mathbf{H}, M) \;P(\mathbf{d}_\mathrm{CC} | \mathbf{H}) \nonumber \\
    &P(\mathbf{H} | \boldsymbol\eta)\;P(\Omega_\mathrm{K})\;P(M)\;P(\boldsymbol\eta)
\end{align}
We sample the joint posterior using Hamiltonian Monte Carlo sampling, implemented in \texttt{PyStan}~\citep{pystan}.

\begin{figure}
    \centering
    \includegraphics[width=.48\textwidth]{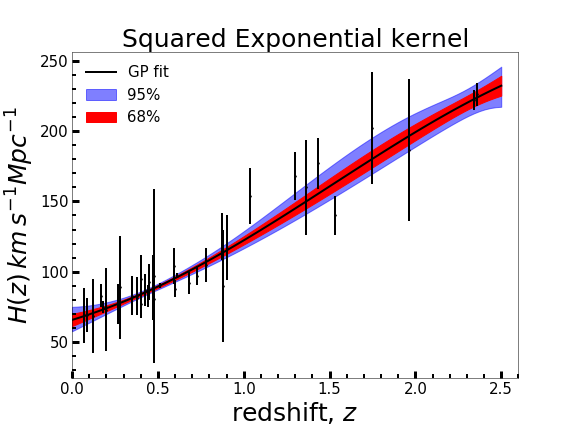}
    \caption{The GP fit (solid black line) and the 68\% and 95\% credible regions (red and blue filled curves) for the reconstructed expansion history, i.e. Hubble parameter as a function of redshift $H(z)$, using the Squared Exponential kernel. The black points correspond to the CC data.}
    \label{fig:expansion_history}
\end{figure}

\section{Results}
\label{sec:results}

We discuss the results obtained using the above methodology applied to current SNe~Ia and CC data, and forecasts for future SNe~Ia and BAO data from Stage IV surveys.

\subsection{Current constraints}

\begin{figure}
    \centering
    \includegraphics[width=.48\textwidth, trim = 10 20 0 20]{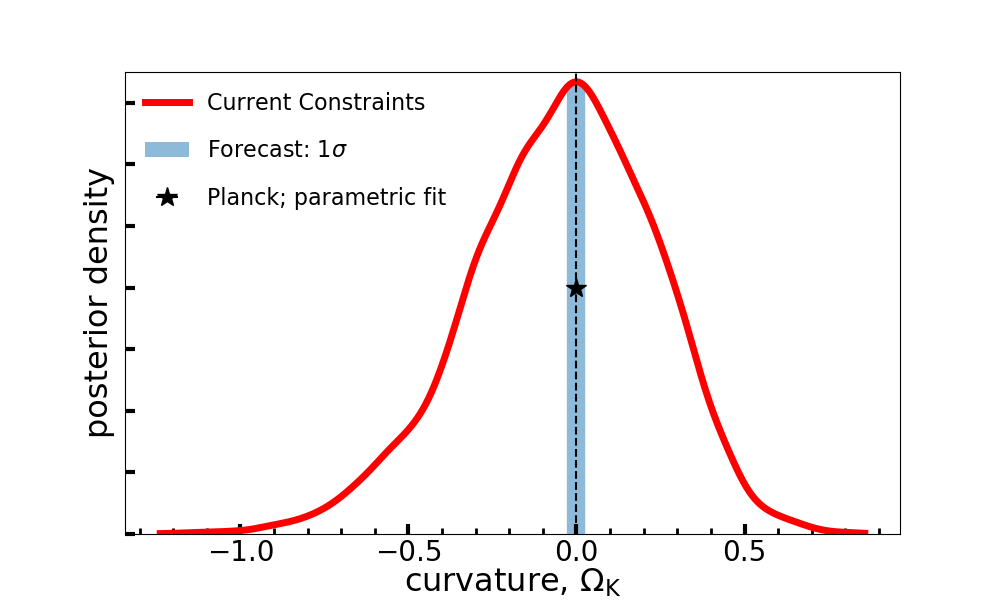}
    \caption{Kernel density estimation (KDE) for $\Omega_{\rm K}$ obtained combining current CC and SNe~Ia data (red line), from which we infer $\Omega_{\rm K} = -0.03 \pm 0.26$. The blue shaded region indicates the expected uncertainty on $\Omega_{\rm K}$ applying our method to future radial BAO and SNe~Ia data, which will allow for an order of magnitude improvement ($\sigma_{\Omega_{\rm K}} \sim 0.026$), comparable to the current constraints from parametric fits to data from the \emph{Planck} satellite (black star; the error bar is the size of the data point, and comparable to the blue shaded region).}
    \label{fig:omegak}
\end{figure}

Applying the methodology described above to the \textit{Pantheon} SNe~Ia and current CC data, we infer $\Omega_{\rm K} = -0.03 \pm 0.26$ (68\% credible region), after marginalizing over all other parameters (including the GP hyper-parameters). Therefore, with current data we are able to reach a precision at the $\sim \mathcal{O}(10^{-1})$ level, approximately one order of magnitude weaker than the parametric constraints from \textit{Planck} primary CMB data alone ($\Omega_{\rm K} = -0.044^{+0.018}_{-0.015}$, under a 7-parameter $\Lambda$CDM+$\Omega_{\rm K}$ model~\citep{Aghanim:2018eyx}). However, we stress that our results rely on late-Universe measurements alone, and do not assume any parametric form for the late-time expansion history (see Figure~\ref{fig:expansion_history} for the GP reconstruction of the Hubble parameter as a function of redshift). Within the achieved precision, we observed no deviations from spatial flatness, as shown in Figure~\ref{fig:omegak}.

We note that other earlier works, including~\cite{Cai:2015pia,Wei:2016xti,Yang:2020bpv,Liu:2020pfa}, derived non-parametric constraints on $\Omega_{\rm K}$ from similar (albeit in some cases older) dataset combinations. Our work is the first to perform a self-consistent joint inference of the cosmological, calibration, and GP hyper-parameters. In the earlier works, the GP regression step of the analysis was instead treated separately from the cosmological parameter inference, which could cause the uncertainties to be underestimated . Some of these earlier works depend on null tests to verify whether the data is consistent with $\Omega_{\rm K}(z)=0$, rather than inferring $\Omega_{\rm K}$ itself. This process requires taking the derivative of noisy data, and is naturally avoided in our methodology. Finally, unlike some of these earlier works, our method does not require knowledge of the absolute $H(z)$ and SNe~Ia calibration, which are naturally marginalized over.

Other studies in the literature have used a cosmographic expansion to describe the Hubble parameter (and related quantities such as the deceleration parameter) in the late universe, and constrain $\Omega_{\rm K}$ using low redshift cosmological probes. For example~\cite{2019PhRvL.123w1101C} adopt strong lensing time-delays and SNe~Ia data to constrain $\Omega_{\rm K}$ at the $\sim 2$-$3 \times 10^{-1}$ level (similar to our constraints). \cite{Jesus:2019jvk} adopts a similar dataset combination, using a cosmokinetic parametrization of the Hubble parameter, deceleration parameter, and comoving distance as a function of redshift, and finds constraints on $\Omega_{\rm K}$ at the $\sim 2$-$3 \times 10^{-1}$ level. Compared to these earlier works, our method makes fewer assumptions on the late-time expansion history, while finding a similar precision on the inferred value of $\Omega_{\rm K}$.

\subsection{Forecasts}

A number of planned Stage IV missions will start taking data in the coming decade. These surveys will make use of multiple, complementary probes, such as BAO, SNe~Ia, and cosmic shear. Since they are all calibrated to the same value of the sound horizon $r_s$, even without knowing $r_s$ (which would require making assumptions about pre-recombination physics), radial BAO measurements probe the evolution of the dimensionless Hubble rate, much as CC data does.~\footnote{This assumes of course that the BAO scale does not evolve at late times, as could occur in certain theories of modified gravity, such as theories which fall within the Horndeski class~\citep[see e.g.][]{Bellini:2015oua}.} Since mock future CC data are not yet available, we forecast the ability of our method to constrain $\Omega_{\rm K}$ from a combination of future SNe~Ia and radial BAO measurements.

Surveys such as the Vera C. Rubin Observatory and the Nancy Grace Roman Space Telescope (NGRST) are expected to increase the yield of high-$z$ SNe~Ia by an order of magnitude compared to current samples, with an improved control over systematic uncertainties. Concerning BAO measurements, the Dark Energy Spectroscopic Instrument~\citep[DESI;][]{2016arXiv161100036D}, Euclid~\citep{2011arXiv1110.3193L}, and NGRST~\citep{2012arXiv1208.4012G} are designed to extend the redshift range of BAO measurements to $z \gg 1$. This will result in significant overlap with the redshift range probed by the SNe~Ia magnitude-redshift relation, with considerable benefits for our method.

We use the forecasts provided for the NGRST SNe~Ia magnitude-redshift relation by~\citet{2018ApJ...867...23H}. The inputs for our analysis include the binned redshift distribution along with the statistical errors on the distance and the systematics covariance matrix, in a similar way to the \textit{Pantheon} compilation described above. Concerning future probes of $E(z)$, we consider mock radial BAO data in the redshift range $0.1 \lesssim z \lesssim 2.0$, matching the expected sensitivity and instrumental specifications of DESI~\citep{2016arXiv161100036D}, Euclid~\citep{2010arXiv1001.0061R}, and NGRST~\citep{2012arXiv1208.4012G}. Mock data to forecast the constraining power of our method is generated assuming a fiducial $\Lambda$CDM cosmology with $\Omega_{\rm M}=0.3$ and $\Omega_{\rm K}=0$.

We find that our method in combination with future SNe~Ia and BAO data considered above will be able to constrain $\Omega_{\rm K}$ with a $1\sigma$ uncertainty of $0.026$ -- an order of magnitude improvement over current \textit{Pantheon} SNe~Ia and CC data (blue shaded area in Figure~\ref{fig:omegak}). This uncertainty is competitive with the current uncertainty coming from \textit{Planck} primary CMB data alone. Therefore, these future measurements will be able to provide a useful and independent test of whether the apparent preference for a closed Universe from \textit{Planck} primary CMB data alone is ``real'', or a statistical fluctuation (as suggested for instance in~\cite{Efstathiou:2019mdh}). We verified that our results are insensitive to the overall calibration of either the SNe~Ia or CC data by re-scaling those data by (different) arbitrary constants, obtaining identical marginal constraints on $\Omega_{\rm K}$.

\section{Discussion and Conclusion}
\label{sec:conclusions}

In this \textit{Letter}, we have presented a novel, self-consistent, non-parametric approach for constraining spatial curvature from late-time cosmological probes alone. After marginalizing over the calibration parameters, GP hyper-parameters, and expansion history $H(z)$ at intermediate redshifts, we infer a value of $\Omega_{\rm K} = -0.03 \pm 0.26$ from current data. This result is consistent with spatial flatness, albeit with an error bar too large to weigh in on the apparent preference for a closed Universe from \textit{Planck} primary CMB data alone. Applying our method to BAO and SNe~Ia data matching the expected sensitivity of upcoming Stage IV missions, we forecast an order of magnitude improvement in constraints on $\Omega_{\rm K}$, which will reach a $\sim \mathcal{O}(10^{-2})$ precision (competitive with the current \textit{Planck} primary CMB constraints).

Our method does not depend on the absolute calibration of $H(z)$, and is thus immune to concerns pertaining, for instance, to the Hubble tension (to the extent that those issues are related to absolute calibration). Moreover, relying exclusively on late-Universe probes, our approach is completely independent of the CMB constraints on $\Omega_{\rm K}$. While parametrized approaches towards constraining spatial curvature will likely remain the standard, our method provides an important complement to parametric constraints.

\section*{Acknowledgements}

S.D. and S.V. are supported by the Isaac Newton Trust and the Kavli Foundation through Newton-Kavli Fellowships. S.D. acknowledges a research fellowship at Lucy Cavendish College. J.A. was supported by the research project grant \emph{Fundamental Physics from Cosmological Surveys} funded by the Swedish Research Council (VR) under Dnr 2017-04212. S.V. acknowledges a College Research Associateship at Homerton College, University of Cambridge. We thank Tom Collett and Till Hoffmann for interesting discussions.

\section*{Data availability}

The data underlying this article will be shared upon request to the corresponding author(s). The associated repository is: \url{https://github.com/sdhawan21/Curvature_GP_LateTime}.

\bibliographystyle{mnras}
\bibliography{Curvature_CC_SNe.bib}

\bsp
\label{lastpage}
\end{document}